# A Novel Paper Recommendation Method Empowered by Knowledge Graph: for Research Beginners

Bangchao Wang[1,2], Ziyang Weng[3*], Yanping Wang[3]
[1]School of Mathematics and Computer Science, Wuhan Textile University, Wuhan, China
[2] School of Computer Science, Wuhan University, Wuhan, China
[3] School of Information Management, Wuhan University, Wuhan, China

***Abstract*—Searching for papers from different academic databases is the most commonly used method by research beginners to obtain cross-domain technical solutions. However, it is usually inefficient and sometimes even useless because traditional search methods neither consider knowledge heterogeneity in different domains nor build the bottom layer of search, including but not limited to the characteristic description text of target solutions and solutions to be excluded. To alleviate this problem, a novel paper recommendation method is proposed herein by introducing "master–slave" domain knowledge graphs, which not only help users express their requirements more accurately but also helps the recommendation system better express knowledge. Specifically, it is not restricted by the cold start problem and is a challenge-oriented method. To identify the rationality and usefulness of the proposed method, we selected two cross-domains and three different academic databases for verification. The experimental results demonstrate the feasibility of obtaining new technical papers in the cross-domain scenario by research beginners using the proposed method. Further, a new research paradigm for research beginners in the early stages is proposed herein.**

## I. Introduction

With the development of artificial intelligence and deep learning, digital resource systems have been garnering remarkable attention. Moreover, considerable amount of new knowledge is mostly captured in digital form and stored in various digital resource systems. Therefore, searching for academic papers in various digital resource libraries has become the main method for researchers to discover potential technical solutions. However, they usually find a large number of academic papers that meet their search criteria; however, obtaining valuable papers that meet their real requirements is challenging [1].

Two drawbacks pertaining to traditional search models lead to this phenomenon. First, traditional search models usually do not consider knowledge heterogeneity in cross-domain research, thereby substantially decreasing effectiveness of a search [2]. Second, in traditional models, users cannot construct the bottom layer of the search which includes but not limited to the characteristic description text of target solutions and solutions to be excluded, and the search keywords are unable to accurately and completely express their requirements [2]. In other words, the models can neither clearly express the problems that users want to solve nor can they help the recommender system express cross-domain knowledge effectively.

Research beginners have to read a large number of papers when they begin with academic research. However, they also have to search for more papers in other domains to obtain suitable technical solutions for domain challenges. Therefore, they need an effective paper recommendation method that can address the aforementioned two drawbacks.

The information extracted from a large number of papers that have been read can provide basic knowledge for constructing a knowledge graph (KG). In this paper, we propose a novel paper recommendation method powered by KG. The overall framework of the method is as follows. First, the user determines the cross-domain scenario wherein the solutions to be obtained solve the target challenge, constructs search keywords, and completes the preliminary search in the academic resource libraries to obtain the primary candidate paper lists. Second, the user constructs a lightweight "master–slave" KG mainly based on in-domain and cross-domain papers. The related information from the KG can be regarded as the underlying data to help improve the bottom layer of the paper recommendation. Third, the paper recommendation process is completed based on the predefined recommendation rules. Fourth, users read the candidate papers, refine the results, and discover potential technical solutions from the confirmed papers.

The main contributions of this study are as follows:

1) A novel paper recommendation method powered by KGs was developed. The introduction of KGs not only helps users accurately express their requirements but also enable the system to build the bottom of the recommendation. The experimental results demonstrate that this method is effective.

2) A new research paradigm to assist research beginners is provided to obtain potential technical solutions across domains in the early stages of research. This paradigm encourages researchers to develop a useful habit of continuously extracting useful information while reading a large number of documents at leisure. With continuous accumulation, the KG domain can be gradually constructed. Finally, the method proposed herein can be used by researchers.

The remainder of this paper is organized as follows. Section II provides background information, Section III reports the process of our approach, and Section IV presents the details of the experiment. The experimental results are presented and discussed in Section V. Finally, in Section VI, the conclusions and future work are discussed.

## II. Background

### A. Paper Recommendation

Commonly used paper recommendation methods mainly include content-based and collaborative filtering-based methods. The former [3][4] recommends users with information similar to what the user has browsed or manipulated based on keywords. In other words, it uses keywords to describe information items, and users would use their own user status table to express their preferences. The latter [5][6] makes recommendations based on the similarity between a user and other users. This method requires a large amount of information such as user behavior, activities, and preferences to be collected and analyzed first.

However, the following drawbacks have been identified:1) The content-based method is a typical user-centric recommendation process. This type of method is mostly based on historical information that the user has browsed or manipulated. However, it cannot regard challenges as clues to recommending papers to users [2].

2) The effect of the collaborative filtering method is restricted by the cold start problem, and its results are often non-personalized. In addition, it does not support challenge-oriented recommendations [2].

### B. *Knowledge Graph*

KG technology—techniques used to construct and apply KGs—is an interdisciplinary subject that integrates cognitive computing, knowledge representation and reasoning, information retrieval and extraction, natural language processing and semantic web, data mining, and machine learning.

The construction methods of KG mainly include two approaches, namely top–down and bottom–up. The top–down construction method involves extracting ontology and pattern information first from high-quality data with the help of structured data sources; then, ontology and pattern information are used to guide data extraction at the bottom layer. The bottom–up method refers to extracting the resource model based on some technical means from the collected data and adds it to the knowledge base as knowledge after review [7].

The top–down KG construction method uses a small number of manually defined abstract ontology patterns as the top-level concept of KG to enrich and gather entities and entity relationships and discover new concepts based on these concepts. This type of method is mostly used to build KGs for specific domains, such as Chinese medicine.

Among the KGs built with a bottom–up approach, the most influential are Google's Knowledge Vault and Microsoft's Satori Knowledge Base. Both of them use publicly collected massive web data as data sources and are constructed by automatically extracting resources and enriching and improving the existing knowledge base. The key actions involved in the bottom–up method include entity, relationship, and attribute extractions [7].

The method used to construct KG mainly depends on the scale, degree of structure, quality of the data source, and quality requirements. If the data source is highly structured and the KG to be constructed is of high quality, using the top–down KG construction method is suitable. If the data source is low in structure and large in scale, but a certain percentage of missing data or errors are allowed in the KG, users can use the bottom–up construction method. If the data source part is structured, large-scale, and allows a certain percentage of missing data or errors in the constructed KG, using the top–down and bottom–up methods to construct KG in a combined manner is suitable.

For example, in [8], a top–down and bottom–up KG construction method is adopted; that is, the KG model layer first builds from the top–down method and then extracts the knowledge from the bottom–up method to build the data layer. In this study, a top–down approach was used to construct KG.

### C. *Paragraph Vector (Doc2Vec)*

The paragraph vector (Doc2Vec) model, is an unsupervised framework that learns continuous distributed vector representations for pieces of text. The texts can be of variable length, ranging from sentences to documents.

In this model, vector representation is trained to be useful for predicting words in a paragraph. More precisely, it concatenates the paragraph vector with several word vectors from a paragraph and predicts the following word in the given context. Both word vectors and paragraph vectors are trained using stochastic gradient descent and backpropagation. While paragraph vectors are unique among paragraphs, word vectors are shared. During prediction, the paragraph vectors are inferred by fixing the word vectors and training the new paragraph vector until convergence [9].

Doc2Vec has two types of models, namely the distributed memory of paragraph vectors (PV-DM [9]) and distributed bag of words version of paragraph vector (PV-DBOW [9]). In this study, the former was used to conduct text vectorization.

## III. OUR APPROACH

### A. *Overview*

Figure 1 describes the overall framework of the proposed KG-based paper recommendation method. This method can be divided into four parts.

**1) Submitting the requirements.** Users first determine the target challenge to be solved and the cross-domain scenario used to obtain solutions. Then, they construct search keywords and search relevant academic resource databases, such as ACM, IEEE, and EI. Subsequently, they extract the titles, abstracts, and keywords from these initially selected papers and construct a text set.

**2) Constructing the domain KG.** Users construct a master–slave KG. The construction of the master KG is based on authoritative, high-quality, and highly representative systematic in-domain literature reviews and is supplemented by core knowledge, basic concepts, and knowledge system frameworks. For slave KG, we only use the representative review papers in the cross-domain as the only data source.

**3) Recommending new technical papers.** The paper recommendation method ranks these initially selected papers based on predefined rules. Specifically, the rule is to determine the technology with the greatest similarity to the target technology feature and the least similarity to the technology to be excluded in the collection of related papers.

**4) Refining result.** Users can identify new technologies having the potential to address challenges by reading the papers in the recommended candidate papers.

In the next section, the details of constructing the master–slave KG is presented.

### B. *Paper-driven top–down KG construction method*

Constructing the master KG in-domain and the slave KG cross-domain is an important basis for paper recommendation, because the background knowledge stored in these two KGs would provide the knowledge premise. According to the survey, this paper adopts the review-driven top–down method to construct the master–slave KG.

The following section introduces the construction process of the master KG in its own domain and the slave KG in the cross-domain.

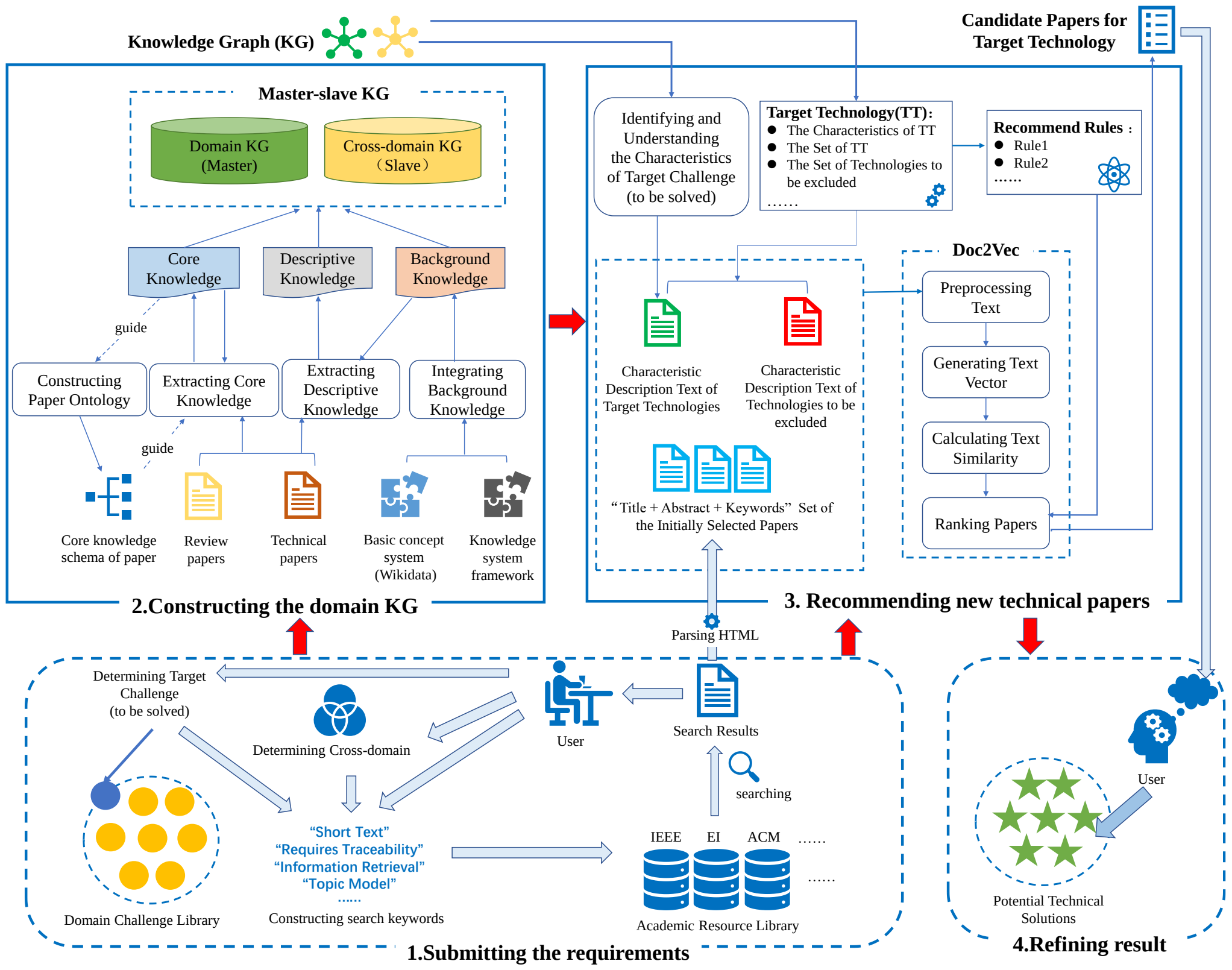

Fig.1. Overview of the paper recommendation method

**1) Construction of the master KG**

The main steps involved the construction of the mater KG is as follows:

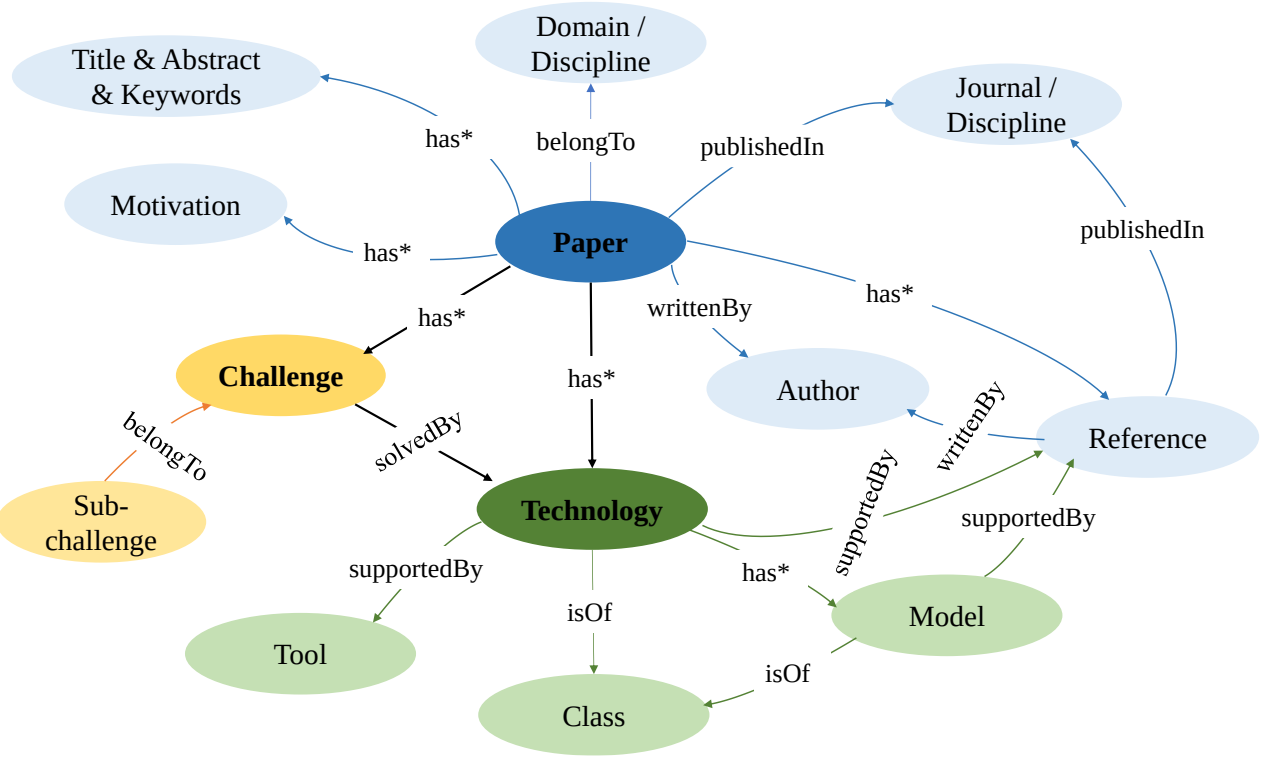

Fig.2. Core knowledge system of the paper (Ontology)

**Step 1: Constructing core knowledge system of paper.**

As shown in Figure 2, the core knowledge system of the paper (hereinafter referred to as paper ontology) is proposed. The construction of this paper ontology was formed after multiple rounds of iterations.

As can be seen from the figure, the paper includes author information (Author), publication information (Journal/Conference), title, abstract, and keywords (Title & Abstract & Keywords), research domain or discipline (Domain/Discipline), research motivation (Motivation), challenge (Challenge), technology (Technology), and reference (Reference).

**Step 2: Extracting data from papers.**

In this step, data are extracted from review papers and technical papers by manually combining them with input tools. The data items included all elements shown in Figure 2. Technical papers were derived from all primary studies related to technology in all selected review papers. During extraction, the paper extraction tool, as shown in Figure 3, was used to help users. This tool greatly reduces the workload of data extraction.

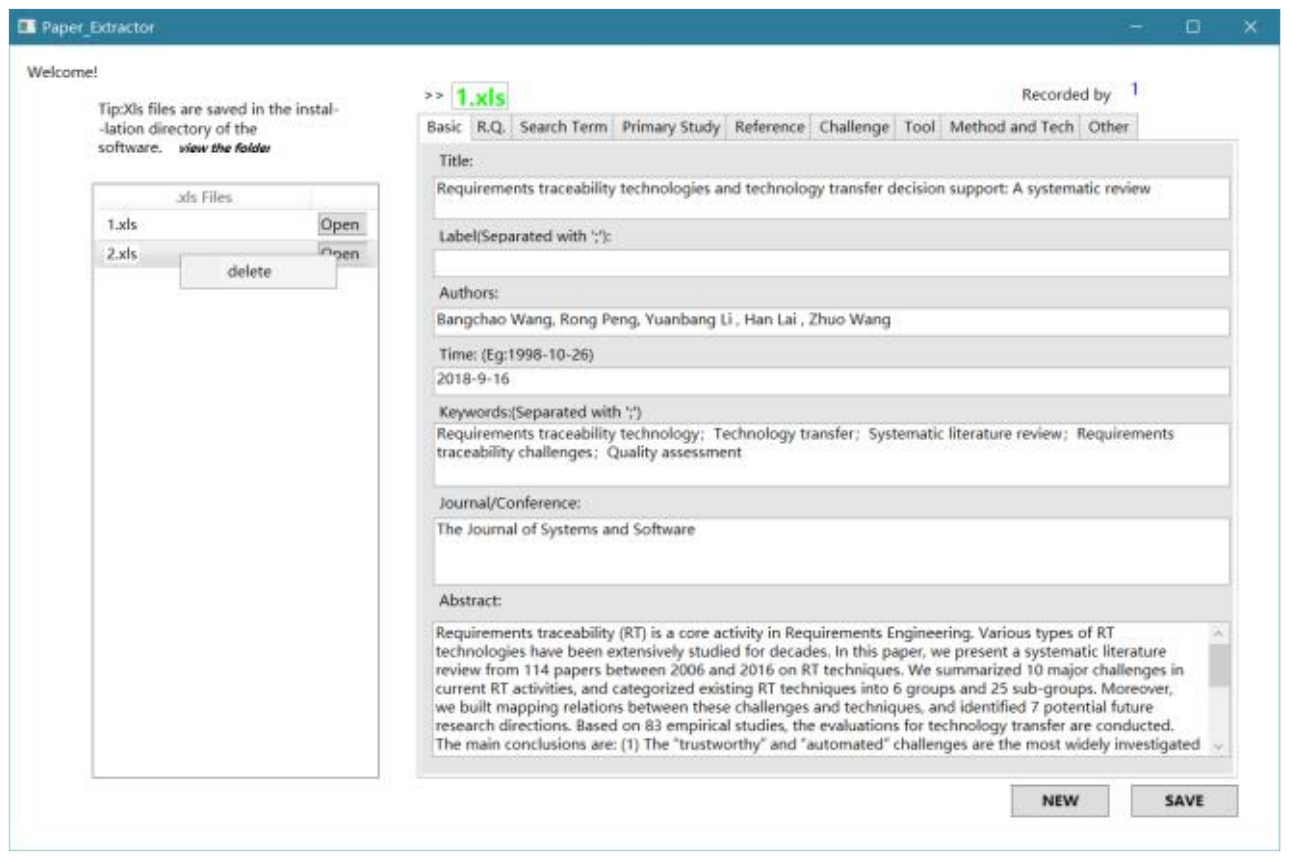

Fig.3. paper extraction tool (Paper_Extractor)

**Step 3: Preprocessing and packaging data.**

The third step is to perform data preprocessing and encapsulation on the data obtained in the previous step. Data preprocessing mainly includes entity alignment, setting null for incomplete data, and data cleaning. For the same entity, knowledge data extracted from different data sources are not the same. Group discussions and negotiations were then used to manually align the entities.

Subsequently, the preprocessed data are encapsulated into the form of comma-separated values (CSV) according to the ontology defined in the first step and imported into the KG for storage. In this method, Neo4J was used to store the data.

**Step 4: Supplementing external knowledge.**

In this step, external knowledge, including core knowledge, basic concepts, and knowledge system frameworks, is supplemented into the master KG. Subsequently, KG is enriched and improved.

**2) Construction of the slave KG**

The construction process of the cross-domain KG is similar to that of the master KG in its own field. The difference is that we only use the representative review papers in the cross-domain as the only data source and then use the entities and entity relationships extracted from the review papers to build the lightweight core KG.

The construction process of the slave KG mainly includes the first three steps of the above four steps. In other words, it does not use review papers as clues dig into the primary studies (technical papers) associated with them and does not use external source knowledge to supplement it.

### C. *Rule-based paper recommendation method*

As shown in Figure 2, the recommended method is based on the characteristic description text of target technologies, characteristic description text of technologies to be excluded and "Title & Abstract & Keywords" set of the initially selected papers. The first two types of texts are obtained from the master–slave KG. In addition, **the characteristic description text of the target challenge identified and understood from KG is also added to the description text of the target technologies**.

The main steps of paper recommendation method is introduced as follows:

**Step 1: Preprocessing text**

In this step, the main preprocessing process contains stop word removal, part-of-speech tagging, and word stemming for the three types of text mentioned above. Subsequently, the metadata obtained after processing were further split into word pairs.

**Step 2: Generating text vector**

In this step, the Doc2Vec algorithm introduced in Section 2.3, is used to fit all the word pairs obtained in the first step to generate embedded text vectors. In particular, the PV-DM model provided by the third-party open-source library gensim[10] was used to implement this task.

**Step 3: Calculating text similarity**

This method needs to calculate the similarity between the two aspects. First, the distance between the initially selected paper and the target technology needs to be calculated, that is, $sim(p, t_{\text{target}})$; second, the distance between the initially selected papers and the technology to be excluded, that is, $sim(p, t_{\text{exclude}})$.

The abstract of a technical paper may describe three aspects, namely the current challenges, the drawbacks of the existing technologies, the characteristics and effects of methods, technologies, or models proposed herein. When there is a description of the drawbacks of the existing technology in the abstract, if it is not distinguished from the newly proposed technology, it would cause serious interference to the similarity measurement of the technical paper with the target technology and the technology to be excluded.

After analysis, it was observed that a drawback description of the existing technology often appears in the first half of the abstract, and its emotional state is usually negative. Therefore, in this study, we constructed the following similarity calculation equations to calculate the similarity between technical paper $p$ and compared technology $t$ ($t_{\text{target}}$ or $t_{\text{exclude}}$).

$$\boldsymbol{sim}(\boldsymbol{p}, \boldsymbol{t}_{\textbf{target}}) = \boldsymbol{cos}(\boldsymbol{p}_1, \boldsymbol{t}_{\textbf{target}}) * \boldsymbol{emo}(\boldsymbol{p}_1) + \boldsymbol{cos}(\boldsymbol{p}_2, \boldsymbol{t}_{\textbf{target}}) \quad \textbf{(1)}$$

$$\boldsymbol{sim}(\boldsymbol{p}, \boldsymbol{t}_{\textbf{exclude}}) = \boldsymbol{cos}(\boldsymbol{p}_1, \boldsymbol{t}_{\textbf{exclude}}) * \boldsymbol{emo}(\boldsymbol{p}_1) + \boldsymbol{cos}(\boldsymbol{p}_2, \boldsymbol{t}_{\textbf{exclude}}) \quad \textbf{(2)}$$

The underlying concept of the two formulas can be divided into the following two aspects:

(1) If the emotional tendency of $p_1$ is negative, the main content of the description is probably the drawback of existing technology. Therefore, the greater the similarity score, the smaller is the similarity between the technical paper $p$ and the compared technology $t$; otherwise, if the emotional tendency is positive, the main content described is the method proposed herein. Then, the greater the similarity score, the greater is the similarity between the technical paper $p$ and the compared technology $t$.

(2) $p_2$ mainly describes the characteristics of the existing technology and the results of the experimental verification. In addition, this is usually unambiguous. Therefore, the greater the similarity score of the technical paper $p$ and the compared technology $t$, the greater is the similarity between them.

In Equation (1), $p$ represents the document vector of the target paper, and we divide it into two parts, namely—$p_1$ and $p_2$. Here, $p_1$ represents the first half of the abstract; $p_2$ represents the second half of the abstract, and the sum of the paper title and keywords; and $t_{\text{target}}$ represents the document vector of the target technology description. $cos$ represents the cosine similarity, $emo$ represents the emotional score, and the score value is automatically obtained by calling the model.prob_classify function in the NLTK platform [11].

The calculation process in Equation (2) is similar to formula 1; however, $t_{\text{exclude}}$ represents the document vector of the technology to be excluded.

**Step 4: Ranking papers**

This step performs two rounds of paper ranking. The ranking process according to the two rules is as follows:

**Rule 1: When $\boldsymbol{sim}(\boldsymbol{p}, \boldsymbol{t}_{\textbf{target}})$ is larger, the paper $\boldsymbol{p}$ is recommended.**

**Rule 2: When $\boldsymbol{sim}(\boldsymbol{p}, \boldsymbol{t}_{\textbf{exclude}})$ is larger, the paper $\boldsymbol{p}$ is excluded.**

Therefore, in the first round, the $sim(p, t_{\text{target}})$ are sorted in descending order to obtain papers that are relatively more related to the target technology. The method delineates the top papers as the scope of the second round of sorting.

In the second round, the $sim(p, t_{\text{exclude}})$, are sorted in ascending order to exclude existing technologies as much as possible. Finally, the papers in top5(Top5) are recommended to users as candidate papers.

Finally, users can discover emerging technologies across domains that have not yet been used in their own research domains, by reading and analyzing the recommended candidate papers.

## IV. Experiment

### A. *Experimental data source*

This experiment uses "software requirements traceability" as the own research domain (in-domain), "information retrieval" (IR), and "machine learning" (ML) as two cross-domains to comprehensively verify the proposed paper recommendation method. The target challenge in the experiment is "short text and fewer artifacts in agile software development environment." Therefore, the experimental data of this experiment mainly included the following two aspects.

**1) Data source for constructing KG**

Table 1 provides the data sources used to construct the master–slave KG in this experiment. Figure 4 shows the global of master KG. Figure 5 shows the local of master KG with paper-challenge technology as the main line.

TABLE I. DATA SOURCE FOR CONSTRUCTING KG

| Data Source | Master KG | Slave KG1 | Slave KG2 |
|---|---|---|---|
| **Review** | [12-18] | [19-21] | [22-24] |
| **Technical Paper** | 278 papers from [12-18] | Null | Null |
| **External Knowledge** | [25] | Null | Null |

**Note: Slave KG1 represents information retrieval, Slave KG2 represents machine learning.**

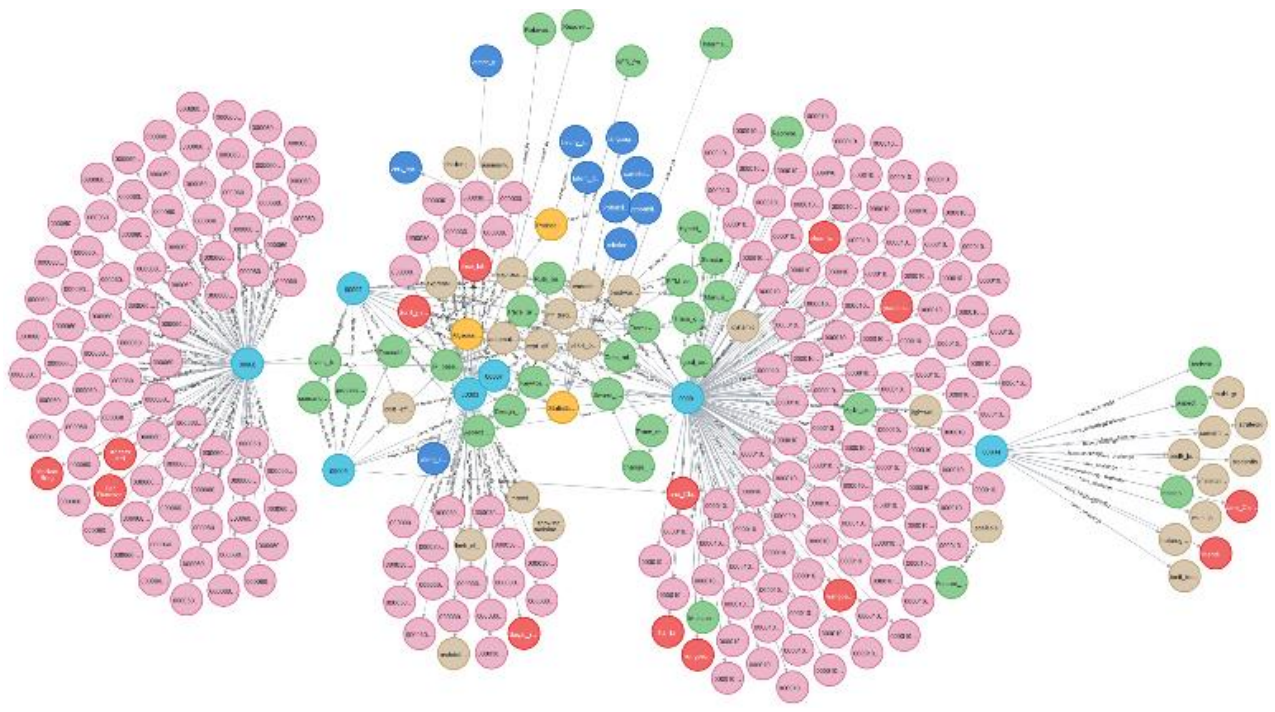

Fig.3. Global of master KG

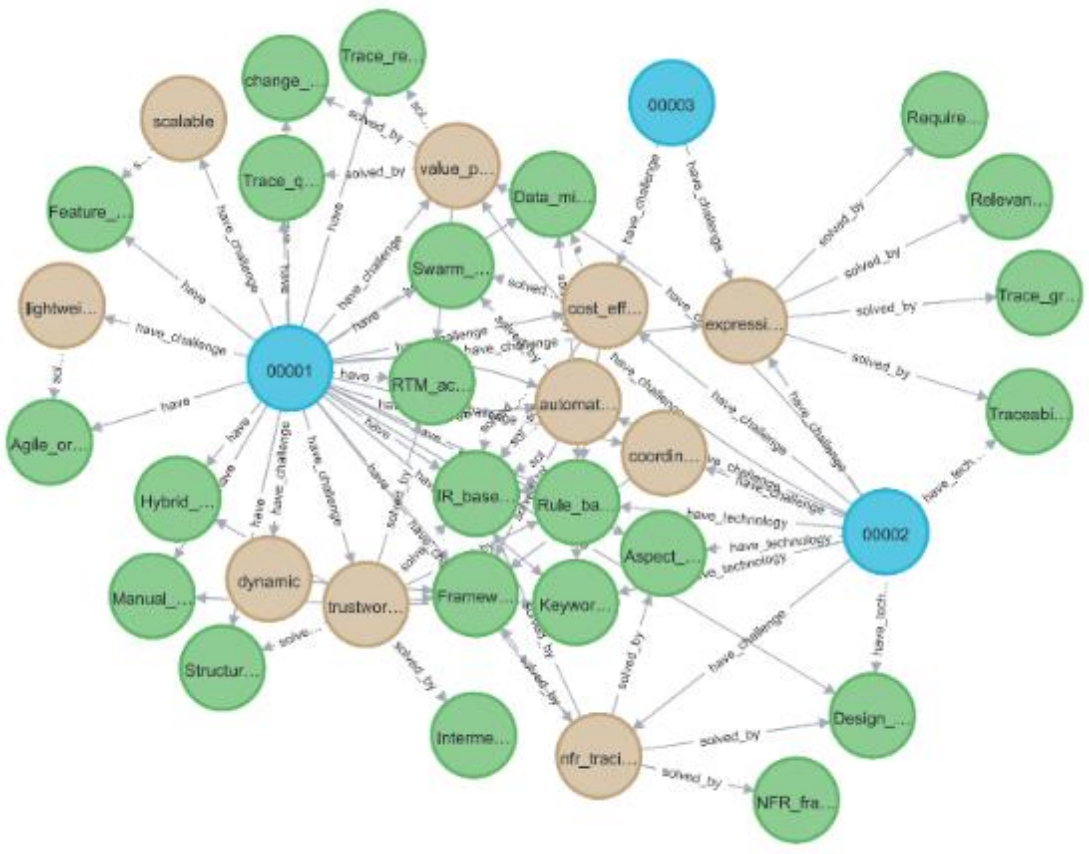

Fig.4. Local of master KG

### 2) Data of initially selected papers in cross-domains

In this experiment, three academic resource databases, Google Scholar, EI, and IEEE, were selected as the data sources. Simultaneously, this experiment formulates information retrieval and machine learning as keywords for searching technical papers.

TABLE II. DATA SOURCE FOR CONSTRUCTING KG

| Data Source | information retrieval | machine learning |
|---|---|---|
| Google Scholar | 100 | 100 |
| EI | 100 | 100 |
| IEEE | 100 | 100 |
| Total | 600 | |

Table 2 gives a brief summary of the experimental data. After performing the search, each database was published from January 2011 to January 2020. The top 100 papers ranked according to relevance are selected, and their titles, abstracts and keyword information are extracted as "Title & Abstract & Keywords" set of the initially selected papers.

## B. *Research questions*

To identify the rationality and usefulness of the proposed paper recommendation method, the following three research questions need to be answered:

**RQ1:** Is the method proposed in this paper effective for cross-domain paper recommendation?

**RQ2:** Is the method proposed in this paper more suitable than the existing methods to help researchers discover potential new cross-domain technical solutions?

**RQ3:** Are there potential new technologies that meet the requirements of the recommended results of this experiment?

## C. *Quality Measures*

Three metrics, namely $TopN$ Effective Rate ($TopN$), $FindN$ Discovery Rate ($FindN$), and Mean Reciprocal Rank ($MRR$), were used to measure the effect of the paper recommendation method proposed herein [2].

$TopN$, as shown in Equation (3), refers to the probability that the target technology can be obtained in the first $N$ recommended papers. According to this definition, $TopN$ is the ratio of the number of papers containing the target technology to the total number of top $N$ papers

$$TopN = \frac{\#\, the\ number\ of\ papers\ that\ include\ object\ technologies}{\#\, the\ number\ of\ TopN\ papers} \quad (3)$$

$FindN$ refers to the number of target technologies that can be obtained in the first $N$ papers. According to the definition, $FindN$ does not need a specific calculation formula, but it judges and counts the total number of target technologies contained in the first $N$ papers by the paper referee team.

$MRR$ is a universal mechanism for evaluating search algorithms. Reciprocal Rank refers to the reciprocal of the ranking of the first correct answer. The equation for the $MRR$ is as follows:

$$MRR = \frac{1}{|Q|}\sum_{i=1}^{|Q|}\frac{1}{rank_i} \quad (4)$$

where $rank_i$ represents the ranking of the first correct answer of the i-th query sentence, and $|Q|$ represents the total number of queries.

When evaluating the effect of the paper recommendation method, it is generally believed that when the $MRR$ value is not less than 0.5, the recommendation system performs well; when the $MRR$ is not less than 0.2, the recommendation system performance is acceptable; when the MRR is less than 0.2, the performance is unacceptable [2]. The $TopN$ also follows this standard.

# V. RESULTS AND DISCUSSION

## A. *Results*

After performing the experiment, in response to the three research questions defined in Section 4.2, the experimental results were summarized and analyzed.

**RQ1: Is the proposed method effective in cross-domain paper recommendation?**

For each keyword, this experiment was performed 20 times on three academic databases. In each experiment, the top 5 (@5, top5) were selected as the recommended results. Therefore, we performed six sets of 120 experiments. Before judging the recommendation result, we established a five-person referee team consisting of one doctoral supervisor, two master supervisors, and two graduate students. In addition, a true set was established.

Table 3 shows the average results for the $TopN$, $FindN$, and $MRR$ of the six groups of experiments.

The average $TopN$ effective rate of all six groups of experiments exceeds 0.2, i.e., at least one of the first five recommended papers can contain the target technology.

From the perspective of the average $TopN$, the effect of the recommended method is acceptable.

TABLE III. AVERAGE RESULTS OF $TopN$, $FindN$, AND $MRR$

| Data Source | Google Scholar | | EI | | IEEE | |
|---|---|---|---|---|---|---|
| | IR | ML | IR | ML | IR | ML |
| Average $TopN$ | 0.8200 | 0.6600 | 0.4200 | 0.7200 | 0.3600 | 0.4200 |
| Average $FindN$ | 4.1000 | 3.3000 | 2.1000 | 3.6000 | 1.8000 | 2.1000 |
| Average $MRR$ | 1.0000 | 0.5263 | 0.2778 | 0.9091 | 0.2632 | 0.4440 |

**Note: IR denotes information retrieval and ML denotes machine learning.**

The average $MRR$ of all six groups of experiments exceeds 0.2, demonstrating that the effect of this recommended method is also acceptable from the perspective of the average $MRR$.

Finally, the average $TopN$, $FindN$, and $MRR$ were combined to evaluate the paper recommendation method. The proposed method is effective for cross-domain paper recommendation.

**RQ2: Is the method proposed in this paper more suitable than the existing methods to help researchers discover potential new cross-domain technical solutions?**

The following is a qualitative analysis and comparison between the paper recommendation method empowered by the KG domain in this paper and the current mainstream academic paper recommendation methods. Table 4 presents the results of this comparison.

TABLE IV. DATA SOURCE FOR CONSTRUCTING KG

| Method | Relate work | ColdStart | Per | User | Challenge |
|---|---|---|---|---|---|
| content-based method | [3-4] | ● | ● | ● | ○ |
| collaborative filtering-based method | [5-6] | ○ | ○ | ○ | ○ |
| Proposed method | This paper | ● | ● | ● | ● |

**Note: (1) ● Positive; (2) ○: Negative.**

From Table 7, does the recommendation method have Cold Start Problem (Cold Start Problem, ColdStart), whether to provide personalized recommendation (Personalized Recommendation, Per), whether it is user-centric (User-centric, User), and whether it supports challenge-oriented (Challenge-oriented, Challenge) are determined as the four perspectives of comparison for this problem. They remarkably affect the ability of technical paper recommendation methods to obtain potential new technical solutions across domains to deal with challenges.

As mentioned in [3][4], the content-based approach cannot deal with challenges and unknowns, and it uses challenges as clues to recommend potential technical papers that may solve specific challenges for users.

It has been reported that the effect of the method based on collaborative filtering is restricted by the cold start problem [5][6]. In addition, the recommendation results are often neither personalized nor user-centric. In addition, it does not support challenge-oriented recommendations.

Compared with these two methods, the method proposed herein is more suitable to help researchers discover potential new cross-domain technical solutions.

**RQ3: Are there potential new technologies that meet the requirements in the recommended results of this experiment?**

By recording and analyzing the recommendation results during this experiment, Yan's two papers published in WWW'2013[26] and AAAI'2015 [27] are the most frequently recommended papers. The main content of these two papers is the word-to-topic model (biterm topic model, BTM). The motivation for proposing the BTM model is to solve the problem of inaccurate topics learned owing to sparse data in a short text environment. This motivation perfectly fits with the challenge that this experiment wants to solve.

Therefore, a recommended technology, BTM, has the potential to meet the requirements of this experiment.

### B. *Validity threats*

In this section, we discuss the potential threats that influence the findings of this study. According to the guidelines for analyzing the validity threats to software engineering methods and processes [28], four aspects are discussed below.

**Conclusion validity**: In our work, we visualized the results and findings to illustrate our improvement. In the future, certain statistical tests may need to be conducted to reduce risk.

**Construct validity**: The evaluation uses three metrics: average $TopN$, average $FindN$, and average $MRR$. They have been widely used to evaluate the effects of paper recommendation methods. Therefore, the metrics used in the experimental evaluation herein can fully quantify our proposed method.

**Internal validity**: The internal validity of our experiment could be affected by the parameter values in the PV-DM model. We chose this value based on our empirical evidence. In the future, we will obtain optimal values by using advanced techniques, such as machine learning.

**External validity**: Different cross-domain and academic databases with various features may lead to different results. To reduce the threats, we selected two cross-domains (IR and ML) and three different academic databases (Google Scholar, EI, IEEE) for verification.

## VI. CONCLUSIONS AND FUTURE WORK

In this paper, a novel paper recommendation method empowered by KG is proposed. By introducing KGs, it can not only help users accurately express their needs but also empower the system to build the bottom of the paper recommendation. The experimental results show that this method is effective and has potential. In addition, a new research paradigm is provided to help research beginners. This encourages researchers to develop a good habit of continuously extracting useful paper information while reading a large number of documents at leisure.

In future, we plan to explore the generality of the proposed method through more cross-domain experimental verification. Subsequently, we will explore how to improve the cost-effectiveness of building a master KG in its own domain.